\documentclass[a4paper]{article}

\usepackage[english]{babel}
\usepackage[utf8x]{inputenc}
\usepackage[T1]{fontenc}
\usepackage{tikz}
\usepackage[]{algorithm2e}
\usepackage{color}
\definecolor{myblue}{rgb}{.8, .8, 1}
\usepackage{marginnote}
\usepackage{amsmath}
\usepackage{empheq}
\usepackage{amsmath,amsfonts,amssymb,amsthm,epsfig,epstopdf,titling,url,array}

\theoremstyle{plain}

\theoremstyle{definition}
\newtheorem{defn}{Definition}[section]

\theoremstyle{remark}

\newlength\mytemplen
\newsavebox\mytempbox

\makeatletter
\newcommand\mybluebox{%
    \@ifnextchar[
       {\@mybluebox}%
       {\@mybluebox[0pt]}}

\def\@mybluebox[#1]{%
    \@ifnextchar[
       {\@@mybluebox[#1]}%
       {\@@mybluebox[#1][0pt]}}

\def\@@mybluebox[#1][#2]#3{
    \sbox\mytempbox{#3}%
    \mytemplen\ht\mytempbox
    \advance\mytemplen #1\relax
    \ht\mytempbox\mytemplen
    \mytemplen\dp\mytempbox
    \advance\mytemplen #2\relax
    \dp\mytempbox\mytemplen
    \colorbox{myblue}{\hspace{1em}\usebox{\mytempbox}\hspace{1em}}}

\makeatother
\usepackage[a4paper,top=3cm,bottom=2cm,left=3cm,right=3cm,marginparwidth=1.75cm]{geometry}

\usepackage{amsmath}
\usepackage{graphicx}
\usepackage[colorinlistoftodos]{todonotes}
\usepackage[colorlinks=true, allcolors=blue]{hyperref}
\pdfoutput=1 

\title{Estimating The Volume Of Breast Tumor On The Digital Breast Tomosynthesis\\
}
\author{Bocar Wane}
\date{March 11, 2017}

\begin{document}
\maketitle

\begin{abstract}
In this paper, we would like to quantitatively measure the tumor volume contained in the breast imaged by the Digital Breast Tomosynthesis (DBT), a reconstructed 3D image. The estimated volume will add to the prognostic value of risk classification of breast cancer. We develop an algorithm that offers an alternative way of estimating the volume of the tumor in a breast. We segment the region of interest by expressing the ratio of tumor region to normal region as a function of the threshold value in the image. Next, we determine the volume of the tumor region as a function of the threshold.  We then find the optimal threshold value that yields the volume of the tumor contained in the breast with the rate of growth of the tumor volume function.  

\end{abstract}

\section{Introduction}

Breast Cancer is the most common cause of death among women \cite{Boyd2009}.
 This paper aims to determine the volume of breast tumor imaged by the Digital Breast Tomosynthesis (DBT) technique. The DBT is a 3D picture constructed from multiple cross sectional 2D images stacked together; and these cross sectional 2D images are obtained from the X- ray beam moving at different angles. 
 Existing methods of measuring the volumetric breast density relay on 3D image modalities like the Magnetic Resonance Imaging (MRI) or attempt to use inference  method from a 2D mamogramms \cite{Jeon2016}. Such methods are either cost prohibited or their accuracy and rigor still remain desirable. In this paper, we particularly propose to estimate the volumetric breast density of a tumor directly from the physical data of the Digital Breast Tomosynthesis. We start by  describing the foundational geometrical laws that govern the breast in the 3D environment, the functional space of which the tumor is subjected to, then we develop an algorithm that segments the tumor in the breast and gives an estimate of the tumor volume contained in the breast.

 \section{Background}
 \subsection{Differentiable Manifolds}
  \begin{defn}
 A differentiable manifold is an abstraction of geometric objects like smooth curves and surfaces in $n$-dimensional space. So a manifold $S$ is a set with a coordinate system. The set $S$ has elements that could be anything from pixel values to probability distributions. The set $S$ must also be endowed with a coordinate system, that is a one-to-one mapping from the elements in $S$ to $R^n$. So each element in $S$ would have a vector representation of $n$ real numbers in $R^n$. Those real number representations of the elements of $S$ are the coordinates of the corresponding element of $S$. So the dimension of $S$ is $n$ and $n = dim S$.\\ So we define a representation of geometric primitive ( elements of $S$) as a mapping from $S$ to a set of numerical parameters \cite{Amari2000}.\\  
 \end{defn}
 
 


 \begin{defn}
  \quad Let $S$ be a set. If there exists a set of coordinate systems $\mathcal{A}$ for $S$ which satisfy the following conditions \cite{DOCARMO1976}:
 \begin{enumerate}
 \item Each coordinate system $\varphi$ of $\mathcal{A}$ is a one-to-one mapping from $S$ to some open subset of $R^n$.
 \item For all $\varphi \in \mathcal{A}$, given an arbitrary one-to-one mapping $\phi$ from $S$ to $R^n$, the following holds:
 \[\phi \in \mathcal{A} \iff \phi \circ \varphi^{-1}\] is a   $C^{\infty}$ diffeomorphism, that is $\phi \circ \varphi^{-1}$ and its inverse $\varphi \circ \phi^{-1}$ are both infinitely many times differentiable or sufficiently smooth.
 
 \end{enumerate}
 \end{defn}
  \begin{defn}
 Let $S$ be a manifold and $\varphi : S \rightarrow R^{n}$ be a coordinate system for $S$. Then for each point p in $S$, $\varphi (p) = [\xi^{1}(p),\cdots, \xi^{n}(p)]$ = $n$ real numbers. Each $\xi^{i}$ is a function from $p \rightarrow \xi^{n}(p)$ and they are the coordinate functions.\\
 Let denote the class of $C^{\infty}$ functions on $S$ by $\mathcal{F}(S)$. For $f, g \in \mathcal{F}(S)$ and $c \in R$, define the following:
 \begin{itemize}
 \item $(f+ g) (p) = f(p) + g(p)$
 \item $(cf) (p) = cf(p)$
 \item $(f.g)(p) = f(p).g(p)$
 \end{itemize}
 The class of $C^{\infty}$ is closed under addition and multiplication. Let $[\xi^{i}]$ and $[\psi^{j}]$ be two coordinate systems, since the coordinate functions are in $C^{\infty}$, the partial derivatives are well defined and the following holds:
 \[    \sum_{j=1}^{n} \frac{\partial \xi^{i}}{\psi^{j}}\frac{\partial \psi^{j}}{\partial \xi^{k}} = \sum_{j=1}^{n} \frac{\partial \psi^{i}}{\partial \xi^{j}}\frac{\partial \xi^{j}}{\partial \psi ^{k}} = \delta_{k}^{i},   \]
 where $\delta_{k}^{i}$ is $1$ if $k = i$, and $0$ otherwise.\\\\ 
 So for any $f \in C^{\infty}$,
 \[ \frac{\partial f}{\partial \psi ^{j}} = \sum_{i = 1}^{n}\frac{\partial \xi^{i}}{\partial \psi^{j}} \frac{\partial f}{\partial \xi^{i}}  \]
 \[  \frac{\partial f}{\partial \xi^{i}} = \sum_{j=1}^{n} \frac{\partial \psi^{j}}{\partial \xi^{i}} \frac{\partial f}{\partial \psi^{j}}     \]
 \end{defn}
 \subsection{Schartz Space}
  \begin{defn}
 The Schartz space $S(R)$ on R is the set of all indefinitely differentiable functions $f$ that are rapidly decreasing at infinity along with all its derivatives so that \[\underset {x\in R}{sup}|x|^{\alpha} |f^{\beta}(x)| < \infty \hspace{0.5cm}  \text{for every} \hspace{0.2cm} \alpha,\beta \geq 0\] 
 The Schartz space $S(R)$ is closed under differentiation and multiplication by polynomials, that is \[ f^{\prime}(x) = \frac{df}{dx} \in S(R) \hspace{0.1cm} \text{and} \hspace{0.1cm} xf(x) \in S(R).  \]
 \end{defn}
 We are considering the natural geometric structure that the surface $S$ inherits from its ambient space, $R^3$ such as the inner product $\langle \cdot, \cdot \rangle$ induced on each tangent plane of $S$ at $p$, $T_{p}(S).$ \\
 
 \subsubsection{First Fundamental Form}
  \begin{defn}
 The quadratic form that corresponds to the inner product induced by $R^3$ on $S$ which allows the measurement of lengths of curves, angles of tangents vectors, and areas of regions on the surface $S$ independently of the ambient space $R^3$ is called the first fundamental form of the surface $S$ at a point $p$. It s defined as 
 \[I_{p}: T_{p}(S) \rightarrow R \hspace {0.3cm} \] 
 such that  \[ I_{p} = \langle v, v\rangle_{p} = |v|^{2} \geq 0 \] for any $v \in T_{p}(S)$.
 Every point p on the oriented surface $S$ is mapped to a point $N(p)$ on the unit sphere $S^{2}$ by the Gauss map $N$ defined by \cite{DOCARMO1976} \[N: S \rightarrow S^{2}\] such that \[S^{2} = \{ (x,y, z) \in R^{3}\}\]
 \end{defn}
 \subsubsection{Second Fundamental Form}
 \begin{defn}

 The point $N(p)$ is the normal vector to $S$ at $p$. The differential of the Gauss map, $dN_{p}$ is a linear map from the tangent space of $S$ at $p$, $T_{p}(S)$, to the tangent space of $S^{2}$ at $N(p)$, $T_{N(p)}(S^{2})$. Since $T_{p}(S)$ and $T_{N(p)}(S^{2})$ are two parallel planes, then the differential $dN_{p}$ can be defined as \[ dN_{p}: T_{p}(S) \rightarrow T_{p}(S)\] and it is called the shape operator.\\
 This map is a self-Adjoint linear map, therefore there corresponds a quadratic form in $T_{p}(S)$ known as the second fundamental form of $S$ at p, $II_{p}$.\[II_{p}(v) = -\langle dN_{p}, v\rangle\] for a unit vector $v\in T_{p}(S)$ tangent to a curve passing through $p$. The second fundamental form measures the normal curvature of a curve passing through p with a tangent unit vector $v$. 
 \end{defn}
 
 \subsubsection{Curvature}
 \begin{defn}
 For a 2-manifold surface in $R^3$,  the local bending is characterized
by its curvature.  The normal curvature $k^{N}(\theta)$ is defined as
the curvature of a curve that belongs to both the surface that the
curve is on and the plane that contains a unit tangent vector
$e_{\theta}$ and the normal vector orthogonal to the surface (locally
approximated by the its tangent plane).  The mean curvature $k_{h}$ is
the average of all normal curvatures that is the normal curvature for
every unit direction $e_{\theta}$ in the tangent plane which would
describe a circle and is given by
\[k_{H} = \frac{1}{2\pi}\int_{0}^{2\pi} k^{N} (\theta) d\theta \]
The principals curvatures $k_{1}$, $k_{2}$ are respectively the
minimum and maximum  curvatures for all normal curvatures.  Therefore,
at any point $p$ on the surface $ k_{H} > min (k_{1}, k_{2}) (p)$. \\
The Gaussian curvature $k_G$ is defined as the product of the extremum
of all the normal curvatures. \[k_{G} = k_{1}k_{2}\]
\end{defn}
\section{Differential forms}

\begin{defn}

A differential form is very similar to vector field such that $ F\vec{i} + G\vec{j} + H\vec{k}$ corresponds to $Fdx + Gdy + Hdz$ where $\vec{i}, \vec{j}, \vec{z}$ are the standard unit vectors on the $x-y-z$ axis. \\\\
A differential k-form can be integrated over an oriented manifold of dimension k such that a differential 1-form quantifies an infinitesimal unit of length, a 2-form an infinitesimal unit of area and a 3-form an infinitesimal unit of volume.\\ 
Performing integration on differential forms offers the advantage of being coordinate independent. With integration being of 3 kinds: \\\\ 
\end{defn}

\begin{itemize}
\item The indefinite integral $\int f$  or the antiderivative
\item the unsigned definite integral $\int _{[a,b]} f(x) dx$ for finding area under a curve
\item  the signed definite integral$\int_{b}^{a} f(x) dx$ for the quantification of a work needed to move an object from a to b.
\end{itemize}
We will use the signed definite integral for  the integration of forms with the following operational properties:\\

\begin{itemize}
\item Two forms $\omega$ and $\nu$ can be added $(\omega + \nu)$ \\
 \item Two forms $\omega$ and $\nu$ can be operated by the  wedge product ($\omega \wedge \nu$) 
 \item A form $\omega$ can be operated by the derivative operator $d$  to get a new form $d\omega$.\\    
\end{itemize}

Note the antisymmetry property of the wedge product: $dx\wedge dy = - dy\wedge dx$ and that for a differential k-form $\omega = fdx^{\alpha}$, $d\omega$ yields a $(k+1) form$ and it is defined by \[d\omega = \sum_{i=1}^{n}\frac{\partial f}{\partial x^{i}} dx^{i}\wedge dx^{\alpha}.\] Moreover the operator d is nipolent that is $d(d\omega) = 0$ \\\\  

The fundamental relationship between the exterior derivative and integration is expressed in Stoke's theorem. The boudary operation $\partial$ maps a $k$ dimensional object into a $k-1$ dimensional object.\\

$\mathbf{Theorem}$: if  $\omega$ is an $(k-1)$ form with compact support on $S$ and $\partial S$ is the boundary of $S$ with its induced orientation, then \[\int_{S}d\omega = \oint_{\partial S} \omega.\] 
\paragraph{Proof:} Let $\phi_{i}$ be a partition on $S$ such that $\sum_{i} \phi_i = 1$
\begin{align}
\oint_{\partial S} \omega &= \oint_{\partial S} \sum_{i} \phi_i\omega \\
& = \sum_{i}  \oint_{\partial S}\phi_i \omega \\
&=\sum_{i}\int_{S} d( \phi_i \omega)\\
&=\sum_{i} \int_{S}(d\phi_i\wedge \omega + \phi_{i}d\omega) \\
&=\sum_{i} \int_{S} d\phi_{i} \wedge \omega + \sum_{i}\int_{s} \phi_{i} d\omega\\
& =\int _{S} (\sum_{i}d\phi_{i}) \wedge \omega + \int_{S}(\sum_{i}\phi_{i})d\omega\\
&=\int_{S}d(\sum_{i} \phi_{i}) \wedge \omega + \int_{S} 1d\omega \\
&=\int_{S}d(1)\wedge \omega + \int_{S}d\omega\\
&=\int_{S} 0 \wedge \omega + \int_{S}d\omega \\
&= \int_{S}d\omega\\
\end{align}

 \section{Algorithm}
  A source light, an X-ray, beams through the breast and the resulting light intensity is measured at the other end. The phenomena followed the Beer- Lambert Law and is modeled by
  \begin{equation}
I = I_{0}e^{-\mu x}
\end{equation}
Where $I_{0}$ is the intensity at the source and $\mu$ is the density
of the medium and $x$ the distance traveled by the light \cite{Stein2003}.\\
If the density $\mu$ of the breast is a function of the distance, then
the X-ray going through the breast will describe a line integral of
$\mu$ along $L$, so the model becomes:
\[
\begin{array}{rcl}
I &=& I_{0}e^{-\int_{L} \mu}\\
\frac{I}{I_{0}} &=& e^{-\int_{L} \mu}\\
\log \left( \frac{I}{I_{0}} \right) &=& \log \left (e^{-\int_{L} \mu}\right)\\
X(\mu) (L) &:=& \int_{L} \mu\\
\end{array}
\]
This density function, $f(x) = \mu (x)$, is a function in Schartz space $\mathcal{S}(R^3)$. \[\mathcal{S}(R^3) = \{ f \in C^{\infty}(R^3) : \|f\|_{\alpha,\beta} < \infty,\quad \forall \alpha, \beta \in {\mathcal{Z}}^n_+\} \] In this paper, we will not evaluate or get an expression of this density function as it will require the use of Radon Transform which  is not the way we would  like to calculate the volume of the tumor. Instead, we consider the breast tissue as a set \[S = \{x_{1}, x_{2},\cdots x_{k} \} \subset R^{3}\]

This set $S$ is a differentiable manifold with real numbers representation of its elements, the pixel values. We are interested in finding those properties of $S$ that  remained constant under any local coordinate systems, that is those properties that are invariant under coordinate transformations.  We looked at the first fundamental form of $S$.\\ 

\begin{figure}[!ht]
 \begin{minipage}{.5\linewidth}
 \includegraphics[width =6cm, height = 4cm]{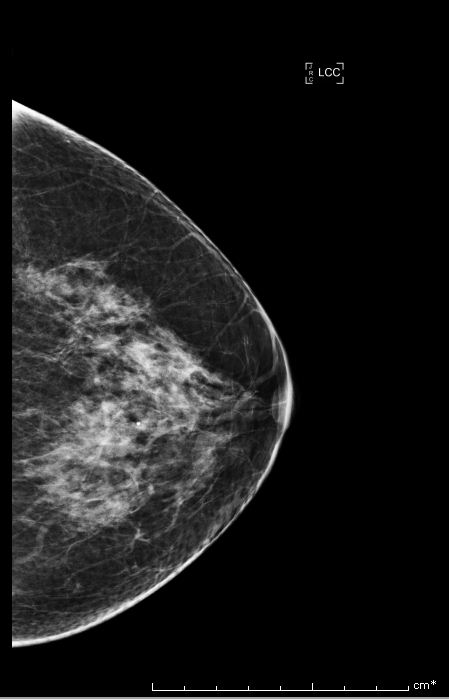} 
 \caption{Bi-rads 4 Breast Image}
 \label{fig: 2}
 {\footnotesize Bi-rads 4 means that the findings on the image are suspicious and that there is an approximately 20 percent to 35 percent chance that a breast cancer is present.\par}
 \end{minipage}
 \begin{minipage}{.5\linewidth}
 
 \[\begin{bmatrix}
 87&208&0&\cdots&117\\
 0&45&186&\cdots&0\\
 \vdots&\vdots&\vdots&\ddots&\vdots\\
 234&0& 56&\cdots&39\\
 \end{bmatrix}\]
 \caption{Image pixel values - $673\times297\times3$}
 \end{minipage}
  \end{figure}

We segment our region of interest from the whole breast image, that is we take a subset of $S$, the tumor region, and  expressed the ratio of tumor tissue to normal tissue,$R$, as a function of the threshold. For a threshold $t$, with $0 \leq t \leq 256$. \\

let \[ X = \{ x_{i} \in S : x_{i} \geq t \}\]
\[ Y = \{ y_{i} \in S : y_{i} < t \}\]

then  $R(t) = \frac{|X|}{|X| + |Y|} $
\\\\ This segmentation is given by the following algorithm:\\
\line(1,0){425}

\begin{algorithm}[H]
 \KwData{Birads 4 Breast Image}
 \KwResult{Segmention of Tumor Area }
 get image\\
 remove zero rows\\
 remove zero columns\\
 let A:= image gradient\\
 do \space  $\sqrt[]{Det A^\top A}$\\
 initialization\;
 k=0\\
 \While{0.2 < Threshold < 1}{
  K:= k +1\;
  index less than Threshold\\
  index more than Threshold\\
  find ratio = number more than threshold/number less than threshold + number more than threshold\\
  Return ratio
 }
 \line(1,0){425}
 \vspace{0.5cm}
 \caption{Tumor Segmentation}
\end{algorithm}

 \begin{figure}[!ht]
 \centering
 \includegraphics[width =8
 cm, height = 7cm]{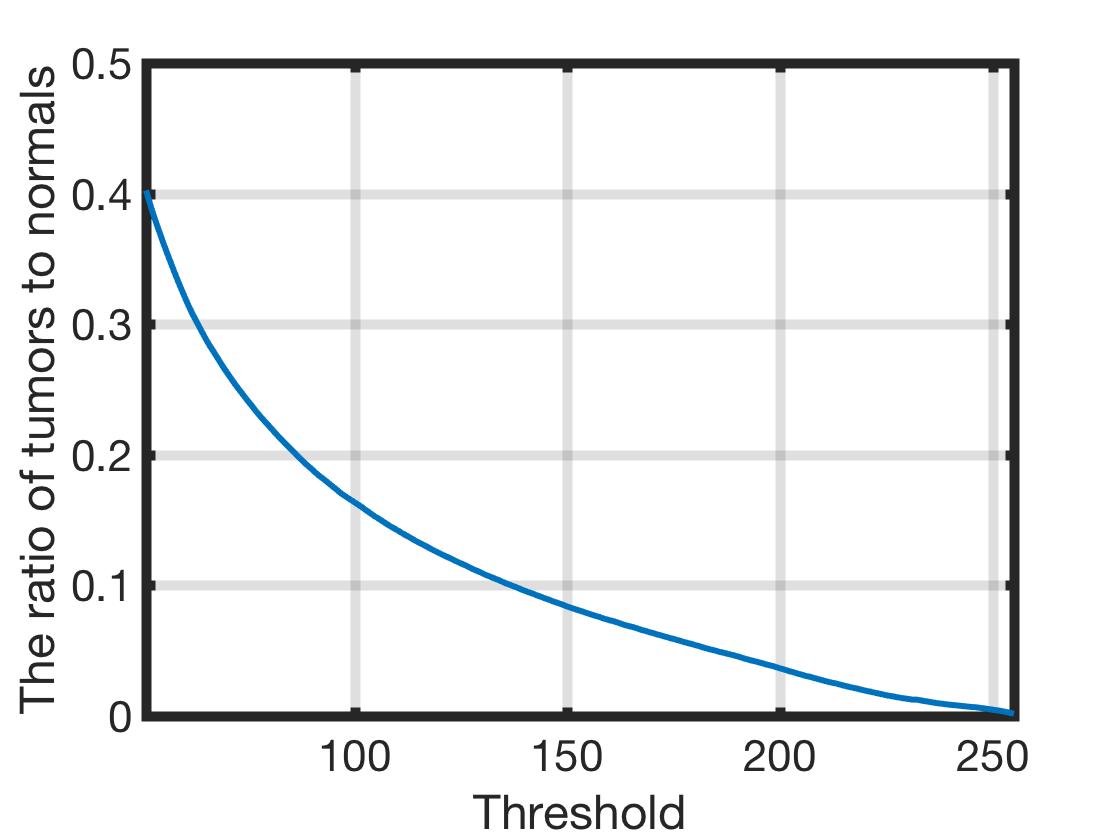}
 \caption{Segmentation}
 \label{fig:ratiotumortonormal.jpg}
 \end{figure}

We proceed to calculate the volume of the  tumor contained in the breast from  the  following Bi-rads 4 breast images.

We take the Gram matrix $G(S)$, a square matrix, of inner products such that \[G(x_{1}, x_{2},\cdots x_{k}) [i,j] = \langle x_{i},x_{j}\rangle \] 
\\The Gramian of $\{x_{1}, x_{2},\cdots x_{k}\}$ is given by the $det \quad G(x_{1}, x_{2},\cdots x_{k}$). So if $\{ x_{1}, x_{2},\cdots x_{k}\}$ are the columms $A \in R^{673\times 297}$ in our image data set, then the \[ det\quad G (S) = det\quad A^{T}A  \] The volume generated by the vectors \[x_{1}, x_{2},\cdots x_{k} \] is given by \[V \{x_{1}, x_{2},\cdots x_{k} \} = \sqrt[]{det \quad G(x_{1}, x_{2},\cdots x_{k}) }\]
\\Given $A \in R^{m\times n}$, the volume of $A$, a first fundamental form, is an intrinsic property of the linear transformation from $R^{3}$ to $R^{2}.$  The volume of the tumor  is then \[R(t)\quad\sqrt[]{det \quad G(x_{1}, x_{2},\cdots x_{k})}.\] 
We express the tumor volume and its growth as a function of the threshold in the following algorithm:\\

\line(1,0){250}

\begin{algorithm}[H]
 \KwData{Tumor Region}
 \KwResult{Tumor volume as a function of the Threshold }
 
 tumor region:= $\sqrt[]{Det A^\top A} \times ratio$ \\
 do derivative of the tumor region\\
 \line(1,0){250} 
 \vspace{0.5cm}
 \caption{Tumor Volume}
\end{algorithm}

 \begin{figure}[!ht]
 \begin{minipage}{.5\linewidth}
 \includegraphics[width =8
 cm, height = 7cm]{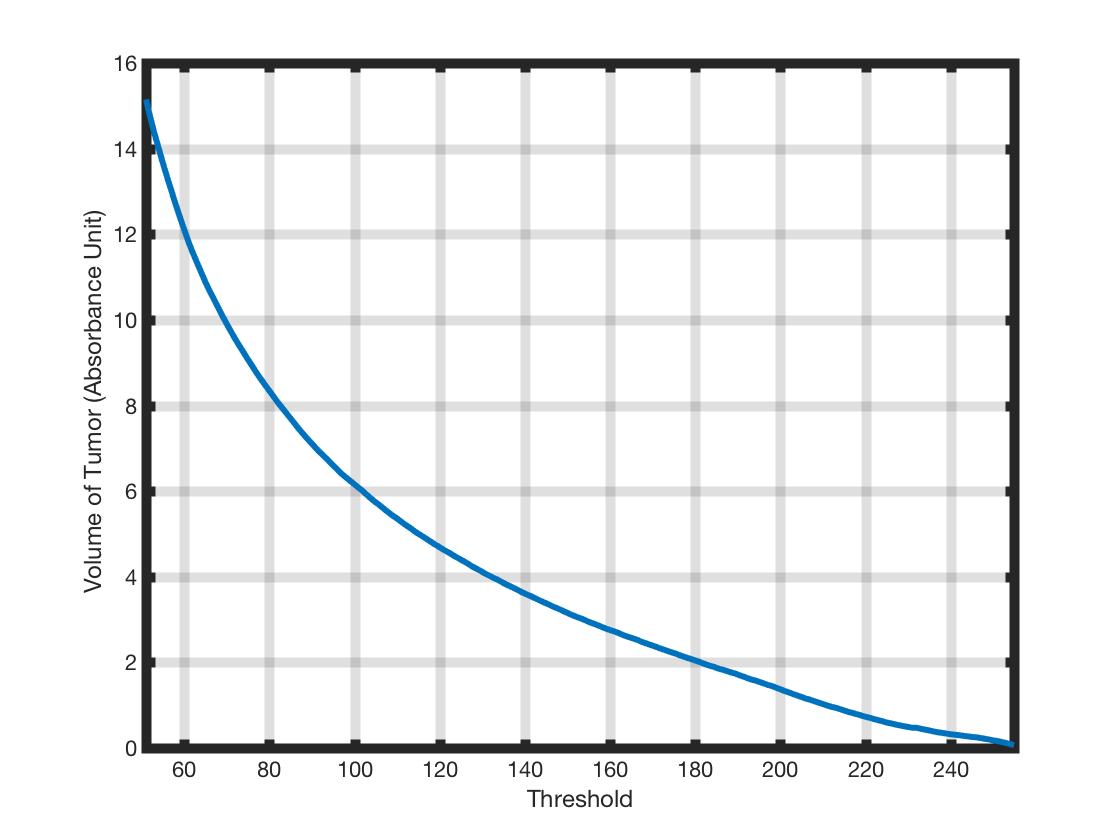}

 \caption{Tumor Volume}
 \label{fig:}
 \end{minipage}
 \begin{minipage}{.5\linewidth}
 \includegraphics[width =8
 cm, height = 7cm]{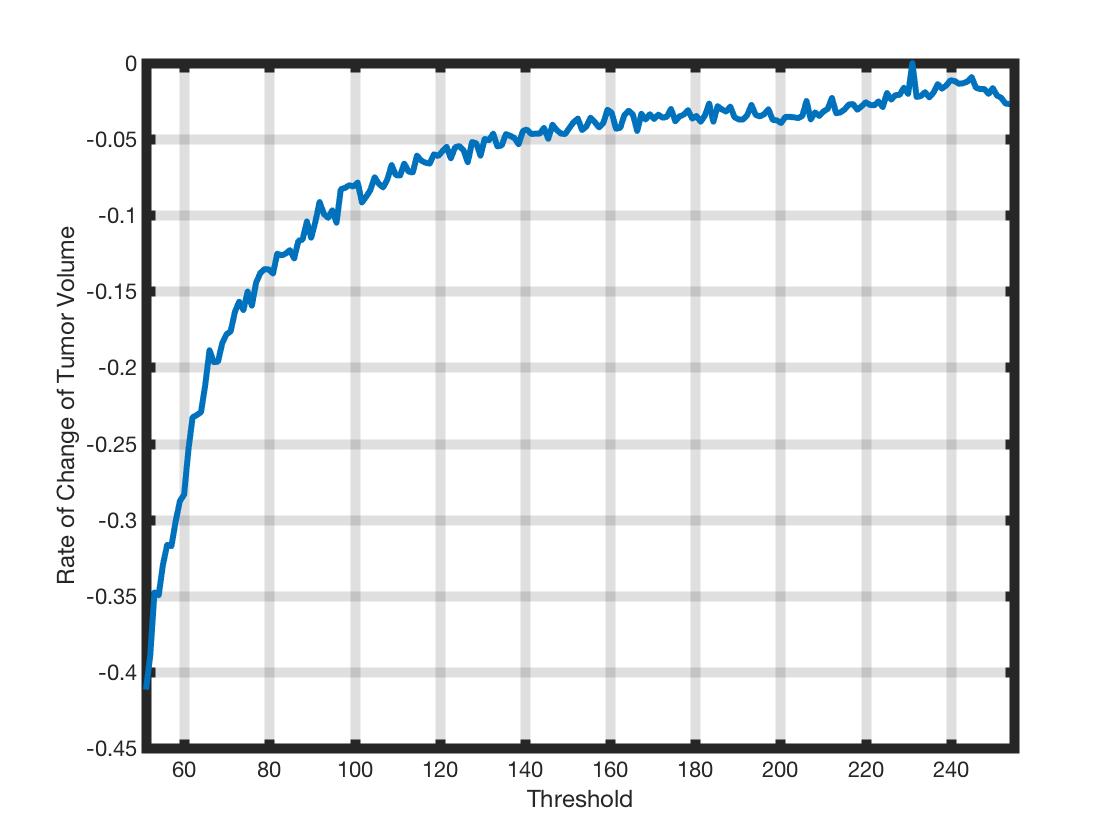}

 \caption{Tumor Growth}
 \label{fig:}
 \end{minipage}
 \end{figure}
 Next, we need to determine the optimal threshold at which we can calculate the volume of the tumor. To do that, we look at the rate of the tumor growth function and determine its inflection points. That is the point or threshold at which the tumor growth function is about to change direction.  We fit a 3rd degree polynomial with the following equation: \[ f(t) = -1.0461\times 10^{-8}t^3 + 5.4201 \times 10^{-6}t^2 
      -0.00090055t + 0.048057 \]
      and solve for the zeros of this function which yields:\\ $ t_{1 } = 119.315\\ t_{2} = 163.912\\ t_{3} = 234.898$\\

\begin{figure}[!ht]
\centering

 \includegraphics[width =8
 cm, height = 7cm]{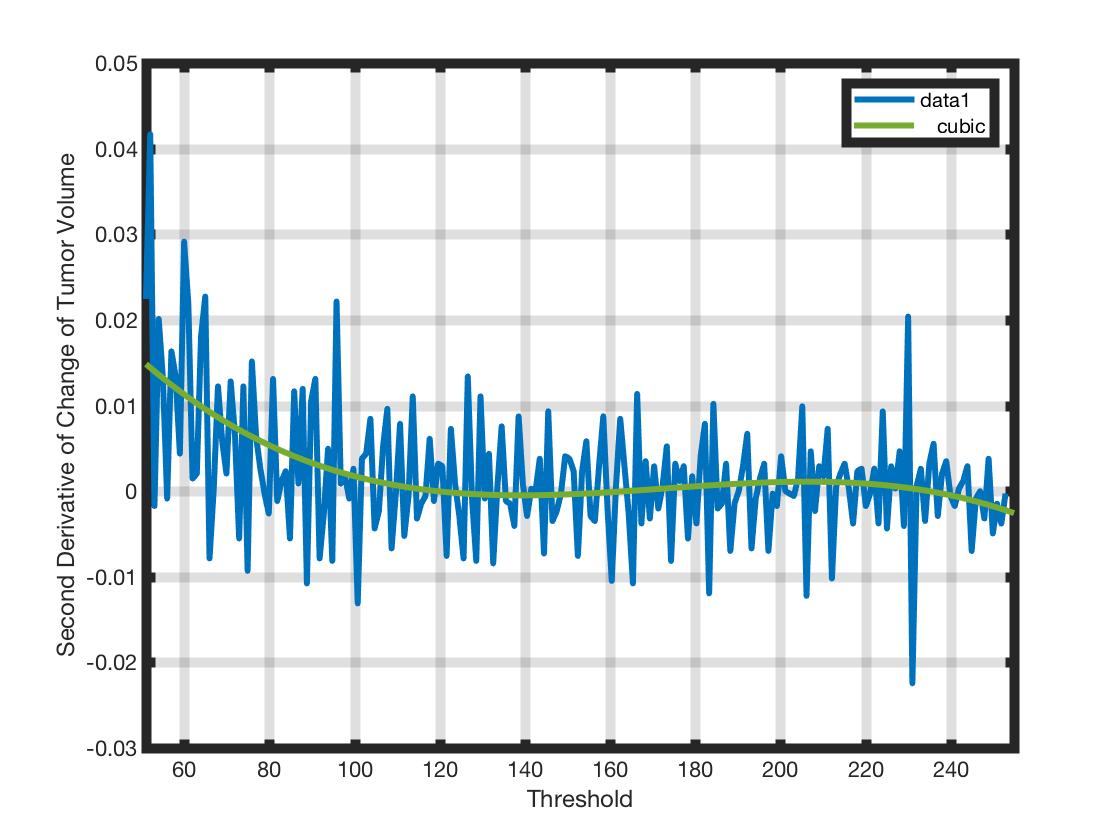}
 \caption{Optimal Threshold}
 \label{fig:}
 \end{figure}

We choose $t = 119.315$ for the optimal threshold upon which the volume of the tumor can be evaluated because this is the first inflection point or the threshold at which the tumor volume function is changing direction for the first time. At this threshold, the absorbance of the X-ray light by the tumor tissue is maximal, and this absorbance is due to the high density of the tumor tissue region or lump. This result in the volume rapidly decreasing since the density and the volume are inversely proportional for a given constant mass.\\\\  We can see from Figure 4 and Figure 5 how this function is rapidly decreasing all the way to $t = 119.315$ and then changes direction to a new inflection point at $t= 163.912$ and continue in a new direction until it hits the last inflection point at $t= 234.898$ and then level off afterward.\\\\
With $t = 119.315$ as the optimal threshold, now we look at the ratio of tumor to normal tissue to determine the percentage of the tumor volume contained in the breast. This volume of the tumor is about $12\%$ of the total breast volume. There are different approaches to measure the total volume of the breast during a breast examination. Of the five methods of measuring the total volume of a breast, we  chose the anatomic method because of its simplicity, accuracy and cost effectiveness \cite{Ragip2011}. The total volume of a breast from the anatomic method is \[breast\quad volume = \frac{\pi}{3}\times MP^2 \times (MR + LR + IR - MP) \]
\begin{itemize}
\item MP = mamary projection
\item MR = medial breast radius
\item LR = lateral breast radius
\item IR = inferior breast radius
\end{itemize}

 \begin{figure}[!ht]
\begin{minipage}{.5\linewidth}
\begin{center}
 \includegraphics[width =5
 cm, height = 2cm]{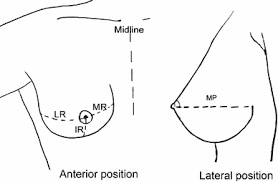}
 \caption{Breast Volume Measurements}
 \end{center}
 \label{fig:}
 \end{minipage}
 \end{figure}
 
 Since our algorithm yields about $12\%$ of the total volume of the breast to be tumorous, then the 
 \begin{empheq}[box={\mybluebox[2pt][2pt]}]{equation*}
    tumor\quad volume = 0.04\pi \times MP^2 \times (MR + LR + IR - MP)
\end{empheq}
\\
\\Similarly, we use the same algorithm to evaluate the tumor volume for 4 Bi-rads 4 images and the results are the following:

\begin{figure}[!ht]
 \begin{minipage}{.5\linewidth}
 \includegraphics[width =6cm, height = 4cm]{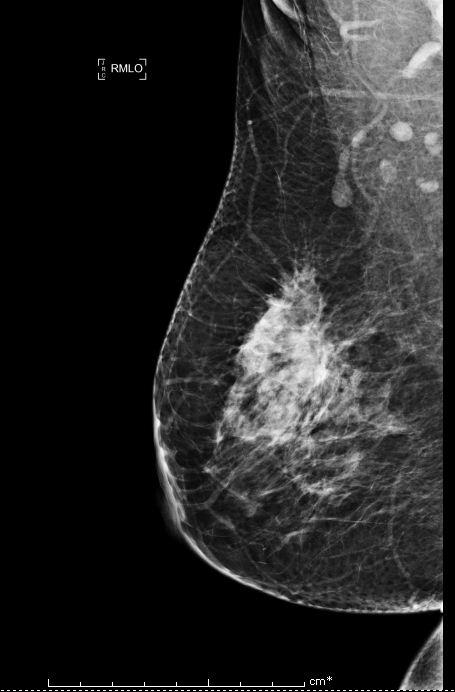} 
 \caption{Bi-rads 4 Breast Image 1}
 \label{fig: 2}
 
 \end{minipage}
 \begin{minipage}{.5\linewidth}
 \includegraphics[width =6cm, height = 4cm]{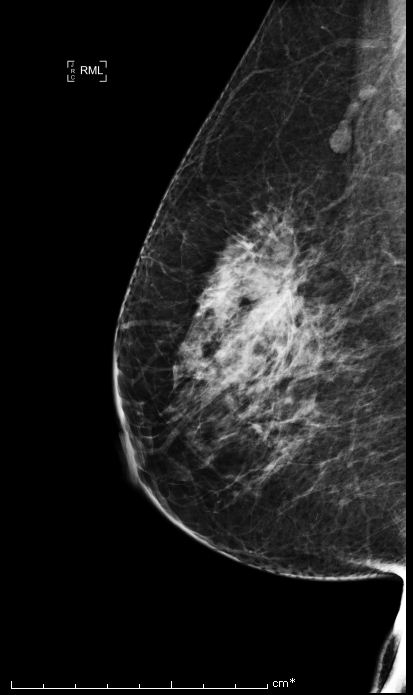}
 
 \caption{Bi-rads 4 Breast Image 2}
 
 \label{fig: 2}
 
 \end{minipage}
  \end{figure}

\begin{figure}[!ht]
 \begin{minipage}{.5\linewidth}
 \includegraphics[width =6cm, height = 4cm]{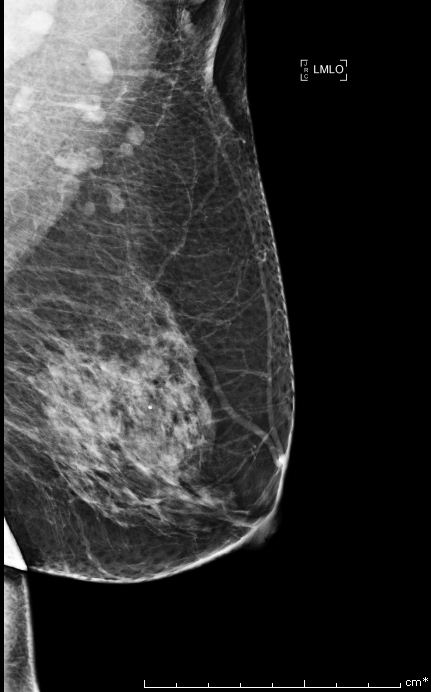} 
 \caption{Bi-rads 4 Breast Image 3}

 \label{fig: 2}
 
 \end{minipage}
 \begin{minipage}{.5\linewidth}
 \includegraphics[width =6cm, height = 4cm]{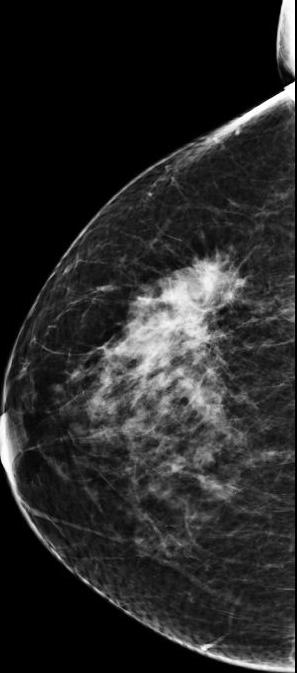} 
 \caption{Bi-rads 4 Breast Image 4}
 \label{fig: 2}
 
 \end{minipage}
  \end{figure}

 \begin{table}[h!]
  \begin{center}
    
 \label{tab:table1}
 \begin{tabular}{l|c|r} 
 \textbf{Bi-rads 4 Image} & \textbf{Optimal Threshold} & \textbf{Tumor Volume}\\
       
      \hline
 Image 1 & 120.574 & 11\% of breast volume\\
  Image 2 & 59.9143 & 45\% of breast volume\\
 Image 3      & 239.696 & 0.24\% of breast volume\\
 Image 4     &  244.590 &0.0287\% of breast volume   \\
    \end{tabular}
  \caption{Tumor Volume for 4 Bi-rads 4 Images}
  \end{center}
  \end{table}
  \newpage
\section{Discussion}
We provide an algorithm that estimates  the volume of the tumor contained in the breast. From the input of the digital tomosynthesis image, we expressed the ratio of tumor tissue to normal tissue as a function of the threshold in order to segment the region  of interest, that is the region of the breast that contained the tumor (Figure:~\ref{fig:ratiotumortonormal.jpg}). We expressed the volume of the tumor as a function of the threshold. We, then, determine the optimal threshold upon which the volume of the tumor can be measured by looking at the rate of the tumor growth function.\\
 The threshold upon which the tumor growth function changes direction for the first time yields the optimal threshold for the rapidly decreasing volume function. Since the X-ray light transmitting through the tumor region is proportional to the density of the tumor region by the Beer-Lambert Law, and that density of that tumor region is inversely proportional to the volume of the tumor region with the mass being constant by the equation- Density = Mass/Volume, then, the optimal threshold is at $t = 119.315$ which yields the volume of the tumor to be about $12\%$ of the breast volume.

\section{Future Work}
We would like to build a Computer Aided Diagnostic (CAD) system, an artificial intelligence, that would classify Bi-rads 4 tomosynthesis breast images as either malignant or benign. We would like to use machine learning on a bigger data set and explore how the proportionality of the size of the tumor volume with respect to the breast volume, along with other features impact the classification the breast tumor as malignant or benign. We will be comparing that experiment results to the biopsy results of the same Bi-rads 4 images in order to evaluate the performance  of the CAD system.  
  




\bibliographystyle{alpha}
\bibliography{sample}

\end{document}